\documentclass[a4paper, 11pt]{article}
\usepackage[english]{babel}
\usepackage{natbib}
\usepackage{hyperref}
\usepackage[hypcap]{caption}
\usepackage[table]{xcolor}
\usepackage[latin1]{inputenc}
\usepackage{amssymb,amsmath}
\usepackage{multirow}
\usepackage{rotating}
\usepackage{subcaption}
\usepackage{graphicx}
\usepackage[hmargin=2.5cm,vmargin=3cm]{geometry}
\usepackage{verbatim}
\usepackage{tikz}
\usetikzlibrary{matrix,shapes,arrows,positioning}
\usepackage{booktabs}
\usetikzlibrary{matrix,shapes,arrows,positioning}
\usepackage{xcolor}
\hypersetup{
    colorlinks,
    linkcolor={red!50!black},
    citecolor={blue!50!black},
    urlcolor={blue!80!black}
}

\allowdisplaybreaks[1]
\usepackage{fancyhdr} 
\usepackage{color, colortbl}
\usepackage{rotating}
\graphicspath{{./Figures/}}
\definecolor{Gray}{gray}{0.9}

\title{Correlated frailty model for analysis of genetic association in family studies}\author{Agnieszka Kr\'{o}l$^1$, Virginie Rondeau$^2$, Yun-Hee Choi$^3$,
Laurent Briollais$^1$}

\date{}
\begin{document}

\maketitle
% Author addresses
\noindent\textit{$^1$Lunenfeld-Tanenbaum Research Institute, Mount Sinai Hospital, Toronto, Canada\\
$^2$Bordeaux Population Health Research Center U1219, Inserm, Biostatistics Team, Bordeaux, France\\
$^3$Department of Epidemiology and Biostatistics, Western University, London, Ontario, Canada
}
\\[2pt]
% E-mail address for correspondence
%\date{\today}
\begin{abstract}
Family-based study designs allow the investigation of gene mutation effects on a disease risk by considering related family members. Some methods have been developed for testing sets of genetic variants in family studies but only very few can handle right-censored time-to-event data. We propose here a correlated frailty model for the analysis of a survival outcome related to cancer in presence of familial correlations. These familial correlations are explained by a residual familial component specified by a kinship matrix and a region- or gene-based specific correlation structure modeled via identical-by-descent (IBD) probability matrix. The proposed approach is  used to  quantify and evaluate the association between a set of common single nucleotide polymorphism (SNPs) or rare variants (or both) from the same genomic region and a survival outcome, e.g. time to disease onset. The model's marginal likelihood is maximized using the Marquardt algorithm. We  evaluated the method by simulations under various scenarios where we varied the family size, the strength of genetic associations from multiple rare variants and the presence or not of residual familial correlation. The results indicate that the correlated frailty model can be valuable in family cancer studies, for example to identify genomic regions significantly associated with the time to cancer onset.
\end{abstract}

\section{Introduction}
\label{sec1}

%family based studies and modeling -> correlated frailty model\\
Family-based studies are attractive and powerful study designs as they allow to investigate the effect of a gene mutation on the disease risk by considering related family members who likely share the disease gene and phenotypes. Thus, in this design, it is crucial to account that individuals from the same family are dependent and ignoring the correlation between the family members can result in invalid inference \citep{Cannon2001}. The growing development of new high-throughput genotyping technologies such as next generation sequencing (NGS) has created new opportunities to study the human genome at a remarkable depth and scale to detect rare-variant association \citep{Lee2014}. Family-based designs have become popular in sequencing studies in recent years, as they allow to establish genetic association in various diseases with samples providing increased power of association detection due to enrichment for multiple copies of rare alleles observed in several relatives \citep{Kazma2011, Jiang2014}.

% testing genetic association in family data \\
A few methods have been developed for testing sets of genetic variants in familial studies by accounting for pedigree relationship. They are usually based on linear or generalized linear mixed-effects models with random effects for the dependence structure or on marginal models with generalized estimating equations (GEE) \citep{Schifano2012,Wang2013}. The association tests are commonly built on integrating kernel machine approach into the regression framework. For instance, for quantitative traits the sequence kernel association test (SKAT) has been extended to famSKAT to account for familial correlation \citep{Chen2013}. For dichotomous phenotypes, \citet{Wang2017} proposed a kernel-based gene set association test using GEE applicable to a mega-analysis. Contrary to the development of association tests for binary and continuous traits, few methods were proposed in the context of right-censored time-to-event data. Rare variants tests were extended to survival outcomes in the framework of kernel based tests and burden tests \citep{Lin2011,Chen2014}. Using functional regression, \citet{Fan2016} developed likelihood ratio test (LRT) for the genetic association with survival outcomes using Cox proportional hazard models. Even less attention have been paid to survival data in family-based studies. \citet{Leclerc2015} proposed an association test with adjustment for familial correlation and used a multiple imputation to treat right-censored observations as missing data. \citet{Chen2015} used the LRT and Wald test in a shared frailty model for survival traits in the family-based setting and proposed an R package RVFam.

%However, the study of the genetic landscape of inherited and acquired mutations in cancer patients could provide invaluable insights into the biological processes involved in complex human diseases, such as cancers, in particular to understand the molecular mechanisms leading to development and progression (Mardis and Wilson, 2009).

In this work, we develop methods for modeling a time-to-event outcome with information on genetic variants through a family relatedness matrix. We use the framework of correlated frailty models that enable to study the intensity process of the event of interest, e.g. disease onset, using random effects for unexplained heterogeneity in the data. We propose a penalized maximum likelihood estimation of the parametric model, in which the penalty is used to correct for a selection bias present in family-based design studies. The assumed source of the heterogeneity are the genetic familial associations that can be represented by e.g. a kinship or identity-by-descent (IBD) probabilities matrix.  The genetic association is quantified by the variance of a frailty term and is evaluated using Wald type test. Depending on a chosen correlation matrix, we test for a residual familial correlation (theoretical kinship matrix) or region-specific genetic familial associations (IBD matrix). We perform a set of simulations studies for the validation of the estimation method of the correlated frailty model and for the empirical assessment of the proposed genetic association test.

%tests available using correlated frailty model -> Wald, LRT, score \\
%what we propose here, objective\\
%outline of the paper (applications)
The proposed method is flexible to various applications of the assessment of genetic association in families sharing rare or common variants and can be used even in the absence of individual genetic information. Depending on the choice of the correlation matrix of familial relatedness, the proposed approach can be used for evaluation of the theoretical or region-specific genetic association. For instance, using a kinship matrix, the proposed method can be applied to a cancer family study to investigate the number of rare and common variants that would explain a familial correlation. Another example would be to evaluate candidate genes using IBD probability matrices in cancer families. Finally, an explanatory genome wide analysis could be performed using both kinship and IBD correlation matrices. 

The rest of this paper is organized as follows. Section 2 describes the correlated frailty model and the estimation methods. In Section 3 we provide a simulation study to evaluate the implemented method for the model estimate and association test following the simulation study structure proposed by \citet{morris2019using}. Finally, Section 4 concludes the paper and discusses the future research.

\section{Methods}
\label{sec2}

\subsection{Correlated frailty model}
We begin by defining a correlated frailty model for a time-to-event family data. For an individual $i\in\{1,\ldots,n_f\}$ from a family $f\in\{1,\ldots,n\}$, let $T_{fi}^*$ denote the time of the event, e.g. disease onset, and $C_{fi}$ the censoring time. We define an indicator $\delta_{fi}=I_{\{T_{fi}^*\leq C_{fi}\}}$ for the event and censoring. For each patient we observe a pair $(T_{fi},\delta_{fi})$, where $T_{fi}=\min(T_{fi}^*,C_{fi})$. We define $p$ nongenetic covariates denoted by $\boldsymbol X_{fi}$ and $\boldsymbol \beta$ is the vector of  regression coefficients. The correlated frailty model is defined as proportional hazard models with frailties:
\begin{align}
\label{mod1}
\lambda_{fi}(t|\boldsymbol b)&=\lambda_{0}(t)\ \textrm{exp}\left(b_{fi}+\boldsymbol X_{fi}^\top\boldsymbol\beta\right) ,\\
\boldsymbol b&=\{b_{fi}, i=1,\ldots,n_f, f=1,\ldots,n\}\sim\mathcal{MVN}(\boldsymbol 0, \boldsymbol\Sigma(\boldsymbol \sigma)), \notag
\end{align}
where $\lambda_0(\cdot)$ is a baseline hazard function, e.g. the Weibull hazard function $\lambda_0(t)= \frac{\rho}{\lambda}(\frac{t}{\lambda})^{\rho-1}$.  The Gaussian random effects, $\boldsymbol b$ are correlated with each other. The dependence structure is defined by the covariance matrix $\boldsymbol \Sigma$, that depends on a vector of parameters $\boldsymbol \sigma$. In practice, the dependence is explained by the genetic associations between members of a family, e.g. the kinship matrix $\boldsymbol K$ and then $\boldsymbol \Sigma(\sigma_1) = \sigma_1^2\times2\times\boldsymbol K$, or the IBD probabilities matrix $D$ and then $\boldsymbol \Sigma(\sigma_2)=\sigma_2^2\times\boldsymbol{D}$. If the interest is in both genetic-specific and residual familial correlation, kinship and IBD matrices can be included in the model, $\boldsymbol \Sigma(\sigma_1, \sigma_2) = \sigma_1^2\times2\times\boldsymbol K+\sigma_2^2\times\boldsymbol{D}$. In all the cases, for identifiability reason, one parameter of the frailty distribution is estimated at a time. If both, the kinship and IBD matrices are used for the correlation structure, one of the parameters $\sigma$ is fixed and the other is estimated. The value of the fixed $\sigma$ can be found by estimating a model with single dependence matrix and use the estimate of $\sigma$ in the final model.

\subsection{Inference and estimation}

In our proposed appraoch, the parameter estimation  is based on maximizing the marginal likelihood obtained from integrating the random effects  over their distribution. The parameters of the model to estimate are $\boldsymbol \Theta=(\lambda,\rho, \sigma, \boldsymbol \beta)^\top$, we assume Weibull baseline hazard functions, but it can be easily generalized to different parametric or semi-parametric, e.g. splines, hazard functions. The marginal likelihood can be written as:
\begin{align}
L(\boldsymbol \Theta)&=\int_{\boldsymbol b}\prod_{f=1}^{n}\prod_{i=1}^{n_f}f_{T_{fi}|\boldsymbol b}(T_{fi},\delta_{fi}|\boldsymbol b;\boldsymbol \Theta)f_{\boldsymbol b}(\boldsymbol b;\boldsymbol \Theta)d\boldsymbol b, \notag
\end{align}
where $f_{T_{fi}|\boldsymbol b}$ is the density of the event and $f_{\boldsymbol b}$ is the density of random effects. As individuals  from different families are independent from each other, the related random effects are independent as well. Noting $\boldsymbol b_{f}=\{b_{fi},i=1,\ldots,n_f\}$ we can simplify the likelihood as follows:
\begin{align*}
L(\boldsymbol \Theta)&=\prod_{f=1}^{n}\int_{b_{f1}}\ldots\int_{b_{fn_f}}\prod_{i=1}^{n_f}f_{T_{fi}|\boldsymbol b_f}(T_{fi},\delta_{fi}|\boldsymbol b_{f};\boldsymbol \Theta)f_{\boldsymbol b_f}(\boldsymbol b_f;\Theta)d\boldsymbol b_f\\
&=\prod_{f=1}^{n}\int_{b_{f1}}\ldots\int_{b_{fn_f}}\prod_{i=1}^{n_f}\left[(\lambda_0e^{b_{fi}+\boldsymbol X_{fi}^\top\boldsymbol \beta})^{\delta_{fi}}e^{-\Lambda_0e^{b_{fi}+\boldsymbol X_{fi}^\top\boldsymbol \beta}}\right]\frac{\boldsymbol b_{f}'\boldsymbol \Sigma(\sigma)^{-1}\boldsymbol b_{f}/2}{(2\pi)^{n_f/2}|\boldsymbol \Sigma(\sigma)|^{1/2}}d\boldsymbol b_f
\end{align*}

The calculation of the likelihood requires to approximate multidimensional integrals with the  dimension equal to the family size. We use a non-adaptive procedure for the Gauss-Hermite quadrature proposed by \citep{Genz1996}, which was designed to estimate multidimensional integrals of the form:
$$I(f)=\frac{1}{(2\pi)^{q/2}}\int_{-\infty}^{+\infty}\int_{-\infty}^{+\infty}\cdots\int_{-\infty}^{+\infty} \text{e}^{-\boldsymbol x^\top \boldsymbol x/2}f(\boldsymbol x)\text{d}x_1\text{d}x_2\cdots\text{d}x_q,$$
where $\boldsymbol x=\{x_i,i=1,\ldots,q\}$ and $f(\cdot)$ the integrand. In our case the integral is with respect to $\boldsymbol b_f\sim\mathcal{N}(\boldsymbol 0,\boldsymbol B)$ ($\boldsymbol B$ is the part of the matrix that corresponds to the family $f$), therefore, a change of variables is required, $\boldsymbol b_f=\boldsymbol x_i\boldsymbol B^{1/2}$. The quadrature rule $Q^{(m,q)}(f)$ is defined as a weighted sum of fully symmetric rules $f[\boldsymbol \lambda_{\boldsymbol p}]$:
$$Q^{(m,q)}(f)=\sum_{\boldsymbol p\in P^{(m,q)}} w_{\boldsymbol p}f[\boldsymbol \lambda_{\boldsymbol p}],$$
where $w_{\boldsymbol p}$ denotes weights, $m$ denotes the degree of a polynomial (it is assumed that $f(\cdot)$ is a polynomial of degree at most $2m+1$), $P^{(m,q)}$ is a set of all distinct $q$-partitions of the integers $0,1,\ldots,m$ and $f[\boldsymbol \lambda_{\boldsymbol p}]$ with $\boldsymbol \lambda_{\boldsymbol p}=\{\lambda_{pi},i=1,\ldots,q\}$ is the fully symmetric sum defined using a set of all permutations of $\boldsymbol p$. This method has the advantage of being faster than the non-adaptive procedure and was found to be efficient, e.g in the \texttt{lcmm} package \citep{Proust2017}. 

The log-likelihood is maximized using the Marquardt algorithm (\citep{Marquardt1963}), a mixture of Newton-Raphson and the steepest descent algorithm. The standard errors are calculated from the diagonal elements of the inverse Hessian matrix of the log-likelihood of the converged model. 

The estimation of the correlated frailty model has been implemented into an R \citep{R} package \texttt{frailtypack} \citep{Rondeau2012,Krol2017}. The extended package is currently available on GitHub (https://github.com/agnieszkakrol/frailtypack) and will be available on CRAN in the future.

\subsection{Consideration of probands and ascertainment correction}

We apply an ascertainment correction using a prospective likelihood \citep{choi2008estimating}. In this method we take into account the ascertainment by dividing the likelihood by the probability of ascertainment, i.e. the probability that the proband was affected before the examination given the proband's genotype, non-genetic covariates and frailty. The penalized log-likelihood $pl^c(\boldsymbol \Theta)$ can be written as:
\begin{equation}
pl^c(\boldsymbol \Theta) = \log(L(\boldsymbol \Theta)) - \text{log}P(T<a_{fp}|\boldsymbol G_{fp},\boldsymbol X_{fp},b_{fp}),
\end{equation}
where $a_{fp}$ is the current age of the proband $p$ from family $f$ and $\boldsymbol G_{fp}$ denotes the genotype.
The penalty can be written as
\begin{align*}
\text{log}P(T<a_{fp}|\boldsymbol G_{fp},\boldsymbol X_{fp},b_{fp})&=\text{log}(1-S(a_{fp}|\boldsymbol G_{fp},\boldsymbol X_{fp},b_{fp})\\
&=\text{log}\left[1-\int_{b_{fp}}\exp\left(-\Lambda(a_{fp}|\boldsymbol G_{fp},\boldsymbol X_{fp},b_{fp})\right)f_{b_{fp}}(b_{fp})d b_{fp}\right],
\end{align*}
where $S(\cdot)$ and $\Lambda(\cdot)$ are the survival and cumulative hazard functions, respectively. We use here the marginal distribution of the random effects for the frailty $b_{fp}$. The variance of the frailty is equal to $\boldsymbol \Sigma(\sigma)_{ii}$ with $i$ being the cumulated indicator of individual $k$ from family $f$ indicating the position of the individual in the complete database.

\subsection{Test for genetic association}

We use one-sided hypothesis tests to evaluate the gene specific dependence via the correlation matrix $\boldsymbol{\Sigma}(\sigma)$. We may consider only one or two association matrices under the alternative model. In the first case, the null model takes form of Cox proportional hazards model and in the second case, the individuals are associated with each other via the matrix that is not tested. In general, we can write the hypotheses:
$$ H_0: \widehat{\sigma^2}= 0 \hspace{0.5cm} vs \hspace{0.5cm} H_1: \widehat{\sigma^2} >0,$$
where $\sigma^2$ represents the estimated parameter in the full model related to the matrix for which we test the association. One method for testing this association is to use the Wald test, in which the test statistic is
$$T_W=\widehat{\sigma^2}/\widehat{Var}(\widehat{\sigma^2}),$$
where $\widehat{Var}(\widehat{\sigma^2})$ is the estimated variance of the frailty parameter obtained from the inverse of the Hessian matrix of the penalized likelihood. Given that the estimated frailty parameter is a constrained estimator, the distribution of the Wald test is a mixture of $\chi^2$ distribution with 0 and 1 degrees of freedom (dof) \citep{Molenberghs2007}.  The Wald test requires estimation of the complete model with the association matrix of interest.

Another method for the evaluation of the genetic association is the likelihood ratio test (LRT), in which the test statistic $T_{LR}=2\log(L(\widehat{\sigma^2})-L(0))$ follows the mixture distribution of $\chi^2$ distribution with 0 and 1 dof. For LRT, the estimation of both complete and restricted models is necessary.

\section{Simulation study}
\label{sec3}
The aim of the simulation study was to assess the quality of inference of the proposed  methodology under different scenarios of family cancer studies. The proposed tests and estimation method were evaluated in a simulation study. The datasets of families (pedigree structure and genotypes) were generated using \texttt{sim1000G} R package \citep{Dimitromanolakis2019} and the time-to-event data was generated with the \texttt{FamEvent} R package \citep{ChoiJSS}.

\subsection{Data-Generating Mechanism}
Firstly, using the simulator \texttt{sim1000G} of genomic regions for related individuals, we generated pedigrees and the associated genotypes. The  genotypic variant data were simulated according to minor allele frequencies (MAF) distribution and linkage disequilibrium (LD) patterns based on a phased Variant Call Format (VCF) file from 1000 genomes Phase III sequencing data \citep{Auton2015} and a genetic map GRCh37 from the corresponding chromosome 4, CEU samples. The VCF file was used to provide the haplotypes for the simulator genotypes. Variants of 100 genes were generated from a range of minimum allele frequency $[0.05;0.1]$. The genotypes for related individuals were simulated according to the meiotic recombination.

We generated 100 and 400 families composed of three generations: parents, one or two children in the second generation and one or two children in the third generation (for each second generation offspring). The correlated frailty model was assumed with gender as the non-genetic covariate (sampled from the Bernoulli distribution with the probability 0.5 for the second generation and third generation) and five dominant genes of the disease. 

We assumed one covariate in the model and fixed the regression coefficient to $\beta=0.5$, it can represent the effect of gender on the risk of developing disease. One of the Weibull parameters, the shape parameter $\rho$ was fixed to $3.0$ and the other, scale parameter $\lambda$ was chosen so that the rate of censored individuals is $20\%$ or $60\%$. The values of $\lambda$ ranged from $58.8$ to $142.9$.  The time of right-censoring was equal to individuals current age sampled from normal distribution with variance 2.5 and mean 95, 75 and 55 for the first, second and third generation, respectively.

The survival data was generated using the modified \textit{simfam} function from the \texttt{FamEvent} package. We simulated time-to-event data using population-based design in which a proband was an affected mutation-carrier. We obtained the data with probands as follows: we used the family structure and genotypes simulated with \texttt{sim1000G}, then for each family we randomly chose a family member for a proband. Finally, we generated times to disease-onset for all individuals until the proband had an affected status, i.e. time to disease was lower than the generated current age of the proband. 

\subsection{Estimand}
The estimands of interest were the parameters $\boldsymbol \theta$ of the correlated frailty model model. The fixed effect $\beta$ and frailty $\sigma$ that capture the effect of observed and unobserved traits on the hazard function were of special interest. Also, we compared empirical power and type I error of the Wald and Likelihood ratio tests.

\subsection{Methods}
In all the scenarios we were testing for the frailty parameter $\sigma_2^2$ related to the IBD probabilities matrix $D=D_1+D_2/2$, where $D_1$ represents the matrix of probabilities of sharing one allele  and $D_2$ of sharing two alleles. In the scenarios with two correlation structures (kinship and IBD), the frailty distribution parameter, $\sigma_1^2$, which corresponded to the kinship matrix, was fixed to 0.5. The frailty parameter for the IBD matrix was assumed 0 for evaluating the type I error and 0.8 for power assessment.

\subsection{Performance Measures}

Performance of the estimation method was evaluated using two metrics: the mean bias of each parameter across 500 replicates and the empirical coverage of its 95\% confidence interval. The power and type I error were measured as the percentage of replicates for which the tested frailty was found statistically significant at the 0.05 $\alpha$ level assuming the respective value of the frailty as described in Section 3.3.

%In the correlated frailty model, we assumed fixed frailty parameter related to the kinship matrix and estimated the frailty parameted for the IBD matrix. Among covariates we included only sex so that the effect of genotypes are captured by the frailty. Then, we tested wheter this parameter is significantly greater than 0.

%Finally, in order to evaluate the proposed estimation method we generated the data with assumed effect of the IBD frailty equal to 1.2. In this scenario we did not assumed any known genotypes as covariates in the survival model. The goal was to evaluate whether we are able to well estimate the model parameters: regression coefficient related to gender, Weibull parameters and frailty variance.  

\subsection{Results}

The results of the estimation of the correlated frailty model with both one and two association structures in all the scenarios, are presented in Table \ref{resSim1}. The regression coefficient $\beta$ was well estimated with the coverage probabilities close to the nominal level 95\%. A small bias was only present in the models with two matrices and censoring rate 20\%. The estimation of the frailty variance $\sigma^2$ was as well satisfying and the bias was only present again in the models with 20\% censoring rate and two correlation matrices. This was related to  smaller average age of the occurring event among the uncensored individuals, close to 40 years, compared to 55 years in the scenarios with higher censoring rates. As consequence, the coverage probabilities related to the frailty parameter in these scenarios were low. Finally, the estimates of the baseline Weibull hazards were unbiased or very little biased. The coverage probabilities were close to 95\% in most of the scenarios except of the scenario with 100 families and 20\% censoring rate. A bias for the parameter $\rho$ was observed in the scenario with 400 families and 20\% censoring rate.

 \begin{table}[]
\begin{center}
\begin{tabular}{lcccccccc}
\hline
              \multicolumn{4}{c}{One correlation matrix}        &   &     \multicolumn{4}{c}{ Two correlation matrices  }   \\
              \multicolumn{4}{c}{mean IBD }                       & &      \multicolumn{4}{c}{Kinship and mean IBD}        \\
\hline
   Parameter          & Mean (SE)                        & ESE  & CP     && Parameter & Mean (SE)                              & ESE  & CP     \\
\hline
             & \multicolumn{8}{c}{ $n = 100$, Censoring rate = 20\%}     \\
$\beta$ = 0.5   & 0.50 (0.13)                      & 0.13 & 95.8\% & &$\beta$ = 0.5   & 0.48 (0.14)                            & 0.13 & 93.2\% \\
$\sigma^2$=0.8  & 0.81 (0.14)                      & 0.13 & 69.9\%& & $\sigma^2$=0.8  &0.58 (0.19)                            & 0.16 & 49.8\% \\
$\rho$=3.0    & 3.01 (0.05)                      & 0.05 & 92.2\% && $\rho$=3.0    & 2.90 (0.05)                            & 0.04 & 84.7\% \\
$\lambda$ = 62.5 & 62.60 (0.13)                     & 0.13 & 94.2\% &&$\lambda$ = 58.8 & 59.05 (0.16)                           & 0.14 & 88.9\% \\
             &\multicolumn{8}{c}{$n = 100$, Censoring  rate = 60\%}        \\
$\beta$ = 0.5   & 0.49 (0.18)                      & 0.18 & 94.8\% & &$\beta$ = 0.5   &0.50 (0.21)                            & 0.20 & 93.4\% \\
$\sigma^2$=0.8  & 0.83 (0.23)                      & 0.23 & 78.3\% & &$\sigma^2$=0.8  & 0.79 (0.34)                            & 0.36 & 86.4\% \\
$\rho$=3.0    & 3.01 (0.06)                      & 0.06 & 95.4\% & &$\rho$=3.0    &3.01 (0.06)                            & 0.06 & 95.6\% \\
$\lambda$= 125  & 124.85 (0.34)                    & 0.34 & 93.2\% & &$\lambda$= 142.9 &142.90 (0.46)                          & 0.44 & 92.6\% \\
             & \multicolumn{8}{c}{$n = 400$, Censoring rate = 20\%}       \\
$\beta$ = 0.5   & 0.50 (0.06)                      & 0.06 & 96.2\% & &$\beta$ = 0.5   &0.47 (0.07)                            & 0.06 & 91.2\% \\
$\sigma^2$=0.8  & 0.76 (0.07)                      & 0.06 & 66.1\% && $\sigma^2$=0.8  &0.48 (0.09)                            & 0.07 & 7.2\%  \\
$\rho$=3.0    & 2.98 (0.02)                      & 0.02 & 94.0\% & &$\rho$=3.0    & 2.85 (0.02)                            & 0.02 & 41.6\% \\
$\lambda$= 62.5 & 62.50 (0.06)                     & 0.06 & 96.8\% & &$\lambda$= 58.8 & 59.03 (0.08)                           & 0.07 & 92.4\% \\
             &\multicolumn{8}{c}{ $n = 400$, Censoring rate =  60\% }     \\
$\beta$= 0.5   & 0.50 (0.09)                      & 0.09 & 93.8\% & &$\beta$= 0.5   & 0.50 (0.10)                            & 0.10 & 95.8\% \\
$\sigma^2$=0.8  & 0.78 (0.11)                      & 0.11 & 74.5\% && $\sigma^2$=0.8  &0.74 (0.14)                            & 0.15 & 77.4\% \\
$\rho$=3.0    & 3.00 (0.03)                      & 0.03 & 93.8\% & &$\rho$=3.0    &2.99 (0.03)                            & 0.03 & 94.0\% \\
$\lambda$ = 125  & 124.76 (0.16)                    & 0.17 & 97.0\% & &$\lambda$ = 142.9 & 142.28 (0.23)                          & 0.22 & 96.6\% \\
\hline
\multicolumn{9}{l}{SE - mean estimated standard error, ESE  - mean empirical standard error, }\\
\multicolumn{9}{l}{CP - coverage probability}
\end{tabular}
\end{center}
\caption{Results of the simulation study using 500 replications and four scenarios for censoring rates and number of families.}
\label{resSim1}
\end{table}

The results of type I error and power of the associations tests using 500 replications are presented in Table \ref{resSim2}. For each scenario we used the Wald test and LRT. For both tests, the type I error and power increase with the number of families analysed and rate of censored observations in the data.

The performance of the association tests depended on the correlation structure. In the simple case, where only one correlation matrix was included, i.e. we did not assume any genetic association via theoretical kinship matrix, the likelihood ratio test performing better in terms of the type I error and power. 

In the case of both correlation structures in the alternative model, only Wald test was available. The type I error was close the nominal level of 5\% only in the scenario with 400 families and 60\% censoring rate. In other scenarios the test was too conservative. The power was over 80\% in the case of the smaller censoring rate and below 70\% otherwise.  The LRT could not be applied to the analysis with kinship and IBD matrices as the likelihoods of the null and alternative models were not comparable. Indeed, in the null hypothesis we included one frailty parameter, related to  the kinship matrix and estimated in the model and in the alternative hypothesis this kinship frailty parameter was fixed and the estimated parameter was related to the IBD matrix.

\begin{table}[]
\begin{center}
\begin{tabular}{lccccc}
\hline
                                        & \multicolumn{2}{c}{\textbf{Type I error}}      & &  \multicolumn{2}{c}{ \textbf{Power}}       \\
                                   &Wald test     & LRT     &  &Wald test        & LRT       \\
\hline
\multicolumn{6}{c}{\textit{No correlation matrix in the null model}}        \\
100 families                            &              &        &                       &            &         \\
\hspace{0.1cm} Censoring rate 20\%                     & 0.8         & 2.6      && 100.0       & 100.0     \\
\hspace{0.1cm } Censoring rate 60\%                     & 1.8          & 4.0     &    & 73.6      &  85.5        \\
400 families                            &              &        &                       &            &    \\
\hspace{0.1cm} Censoring rate 20\%                     & 2        & 4.4            && 100.0        & 100.0   \\
\hspace{0.1cm} Censoring rate 60\%                     & 4.2       & 5.4    && 100.0        & 100    .0  \\
        &              &        &                       &            &     \\
\multicolumn{6}{c}{\textit{Kinship matrix in the null model}}      \\
100 families                            &              &        &                       &            &        \\
\hspace{0.1cm} Censoring rate 20\%                     & 1.8          & -       && 83.3    & -                      \\
\hspace{0.1cm} Censoring rate 60\%                     & 1.9             & -             & & 61.3  & -                      \\
400 families                            &              &        &                       &            &        \\
\hspace{0.1cm} Censoring rate 20\%                     & 2.8         & -                    & &  100.0          & -                 \\
\hspace{0.1cm} Censoring rate 60\%                     & 4.2    & -         & & 69.8        & -                       \\
\hline
\end{tabular}
\caption{Type I error and power of the genetic association tests in the simulated data using 500 replications for each scenario.}
\label{resSim2}
\end{center}
\end{table}

\section{Discussion}
\label{sec5}

In this work, we introduced an approach for modeling time-to-event data with the consideration of the association between family members data. We extended the  shared frailty model that assumes a common unobserved trait within each cluster, represented by the random variable of variance $\sigma^2$ called frailty, by allowing for subject-specific frailty that is correlated to other members of the same family via a matrix of genetic association, such as kinship matrix or IBD matrix. We proposed also an approach for taking into account both kinship and IBD matrices in the analysis of the families time-to-event data. The inference is based on maximizing the marginal likelihood that uses a penalization to account for proband in the data and the maximization is performed using  the Marquardt algorithm.

The simulation study supported the estimation method and the use of the model for testing of genetic association in family studies in terms of power and type I error. When considering one source of correlation, we found unbiased results for all the scenarios. However, in one of the considered scenarios, when the censoring rate was 20\% the coverage probability was much lower than 95\% for the frailty variance. After investigating the generated data, we found that this was a result of data-generating mechanism, as uncensored individuals experienced the event much earlier, on average, than uncensored individuals in the scenario with the higher censoring rate. Undercoverage was even more pronounced when two genetic association matrices were incorporated, and the frailty variance was also underestimated in this setting.

The correlated frailty model proposed in this work is a straightforward method for finding the genetic association for cancer onset in family-based studies since the log-likelihood is maximized in the Marquardt algorithm and the standard error of the frailty parameter is obtained from the Hessian matrix of the log-likelihood. As consequence, the Wald test and Likelihood Ratio Test are directly used to test the significance of the kinship matrix in the cancer risk and to identify candidate
rare sequence variants shared by family members.

Another approach for testing the genetic association using the frailty model, would be to implement a score test following the framework outlined by \citet{Commenges1997}. Details of the score statistics methodology are given in Appendix. This approach will be explored more in the future research.

Beyond the limitation to only Wald and LR tests, our approach can be improved and extended in other aspects as well. The important challenge of the correlated frailty model is the use of Genz procedure as the faster alternative for non-adaptive Gauss-Hermite quadrature to approximate multidimensional integrals. In case of large datasets, the estimation of the model can take several minutes and therefore another integration methods could be considered such as Quasi Monte Carlo integration \citep{pan2007quasi}. In the next step we would like to extend the proposed method to a joint model for recurrent events and time-to-event data in which we incorporate the genetic information via random effects. In the context of colorectal cancer, the recurrent events could represent the process of repeating screening visits of LS family members. This method will enable us to better evaluate the genetic bias in affected families.

In conclusion, we proposed a new approach for correlated frailty models for family-based NGS data. This methodology offers a flexible framework for investigating genetic familial association with survival outcomes. Although developed in the context of cancer studies,it may also be useful for other chronic diseases with variable age at onset and familial clustering. Open-source software is available online to facilitate its application.

\section{Software}
\label{sec6}
All the models were estimated using the R package
frailtypack. The code for the simulation study is available upon request.

\section*{Acknowledgments}
We thank CIHR Stage and CANSSI for funding the postdoctoral research of Agnieszka Kr\'ol. \\

\noindent{\it Conflict of Interest}: None declared.

\bibliographystyle{biorefs}
\bibliography{library}

\section*{Appendix}
The score statistic is the gradient of the likelihood of observations $1,\cdots,N$  ($N$ total number or individuals) under the null hypothesis which in the case of the correlated frailty model with correlation matrix $\boldsymbol A$ representing matrices $\boldsymbol K$ or $\boldsymbol D$ can be written as:
$$ T_S=U^\top\boldsymbol{A}U-tr(\hat{I}_{\boldsymbol b}\boldsymbol{A}),$$
where $U$ is the vector of martingale residuals at $T_i$ under the null hypothesis, $U_{i} = \delta_{i}-R_{i}(T_{i}) = \delta_{i} -\widehat{b_{i}^0} \int_{0}^{T_{i}} \lambda_0(u)\exp(\boldsymbol X_{i}^\top\boldsymbol{\beta})du$, with $\widehat{b_{i}^0}$ the empirical Bayes estimator of the random effects under the null model. The Fischer information matrix under the null hypothesis has the elements:
$$(\hat{I}_{\boldsymbol b})_{ij}=-\frac{\partial^2 (\log L)}{\partial b_{i}\partial b_{j}}(0)$$
The score statistic can be decomposed into a pairwise correlation term (WPC) and an overdispersion term (OD). As the WPC statistic, $T_{WPC}$, was shown to be more robust than the OD term, we will use $T_{WPC}$ denoted here $T$ that can be written as
$$T_S=U^\top(\boldsymbol{A}-I)U-tr(\hat{I}_{\boldsymbol b}(\boldsymbol{A}-1))=U^\top\boldsymbol{A}^*U-tr(\hat{I}_{\boldsymbol b}\boldsymbol{A}^*),$$
where $I$ is the identity matrix and $\boldsymbol A^*=\boldsymbol A -I$. Considering the conditional independence of the likelihood given the random effects, the information matrix $\hat{I}_{\boldsymbol b}$ is diagonal and therefore $tr(\hat{I}_{\boldsymbol b}\boldsymbol{A}^*)=0$. The test statistics follows asymptotically the normal distribution $T_S\sim\mathcal{N}(0, Var(T_S))$ with $Var(T_S)=2\sum_{i=1}^N\sum_{j=1}^Nw^2_{ij}v_iv_j$, where $v_i$ is the cumulative baseline hazard at $T_i$ under the null and $a_{ij}$ is the $ij$-th element of matrix $\boldsymbol A$ being $2\times\boldsymbol K$ in case of the kinship matrix, and $\boldsymbol D$ in case of the IBD probabilities matrix.

\end{document}